\documentclass[a4paper, 10pt]{IEEEconf}      % Use this line for a4
\usepackage{graphics} % for pdf, bitmapped graphics files
\usepackage{epsfig} % for postscript graphics files
\usepackage{mathptmx} % assumes new font selection scheme installed
\usepackage{times} % assumes new font selection scheme installed
\usepackage{amsmath} % assumes amsmath package installed
\usepackage{amssymb}  % assumes amsmath package installed
\usepackage{subcaption}

\title{\LARGE \bf
Design and simulation of 1.28 Tbps dense wavelength division multiplex system suitable for long haul backbone
}

\author{Akinwumi A. Amusan$^{1*}$ and Elizabeth A. Amusan$^{2}$% <-this % stops a space
\thanks{$^{1}$Department of Electrical and Electronics Engineering, Elizade University, Ilara-mokin, Nigeria.
     \newline    {\small akinwumi.amusan$@$elizadeuniversity.edu.ng, akinwummy$@$gmail.com}}%
\thanks{$^{2}$Department of Computer Science and Engineering, Ladoke Akintola University of Technology, Ogbomoso, Nigeria. \newline {\small eaadewusi$@$lautech.edu.ng}}%
}

\begin{document}

\maketitle
\thispagestyle{empty}
\pagestyle{empty}

%%%%%%%%%%%%%%%%%%%%%%%%%%%%%%%%%%%%%%%%%%%%%%%%%%%%%%%%%%%%%%%%%%%%%%%%%%%%%%%%
\begin{abstract}

Wavelength division multiplex (WDM) system with on / off keying (OOK) modulation and direct detection (DD) is generally simple to implement, less expensive and energy efficient. The determination of the possible design capacity limit, in terms of the bit rate-distance product in WDM-OOK-DD systems is therefore crucial, considering transmitter / receiver simplicity, as well as energy and cost efficiency. 
A 32-channel wavelength division multiplex system is designed and simulated over 1000 km fiber length using Optsim commercial simulation software. The standard channel spacing of 0.4 nm was used in the C-band range from 1.5436-1.556 nm. Each channel used the simple non return to zero - on / off keying (NRZ-OOK) modulation format to modulate a continuous wave (CW) laser source at 40 Gbps using an external modulator, while the receiver uses a DD scheme. It is proposed that the design will be suitable for long haul mobile backbone in a national network, since up to 1.28 Tbps data rates can be transmitted over 1000 km. A bit rate-length product of 1.28 Pbps.km was obtained as the optimum capacity limit in 32 channel dispersion managed WDM-OOK-DD system.

\end{abstract}

%%%%%%%%%%%%%%%%%%%%%%%%%%%%%%%%%%%%%%%%%%%%%%%%%%%%%%%%%%%%%%%%%%%%%%%%%%%%%%%%
\section*{Introduction}

The need for fast and reliable exchange of information has increased further in our present society. The reliable operation of industries, businesses and banks; vehicles and transportation systems; household entertainment electronics and the global flow of news and knowledge rely on advanced telecommunication infrastructure. Numerous services and applications of information and communication technology in medical diagnosis and treatment, traffic safety and guidance, as well as, the Internet of things are emerging, stretching the needs for high-capacity communications even further \cite{c1}. By 2020, the mobile communication system will move to 5G, in which the users data rate will be at least 1 Gbps (Giga bits per second), with connection density reaching 1 million connections per square km, end-to-end latency in milliseconds level, traffic volume density of tens of Gbps per square km, mobility greater than 500 km per hour and peak data rate in tens of Gbps \cite{c2}. This implies that the backbone of the mobile and data service providers will require more fiber connections, since fiber optics has the potential to provide the huge bandwidth capacity required for communication in our present society.
In fiber optics systems, light pulses are sent through an optical fiber, and the signals are propagated by total internal reflection between a high refractive index core and low refractive index cladding. Although, fiber systems can meet the huge bandwidth demand, yet the fiber bandwidth needs to be harnessed by modulating into higher data rates before transmission. Dense-wavelength-division-multiplexing (DWDM) and erbium-doped fiber amplifier (EDFA) technologies have been used to increase the capacity by using simple on / off keying (OOK) format and direct detection up to 10 Gbps per channel \cite{c3}. In DWDM, several modulated light carriers with wavelengths spaced at constant channel spacing (often less than 1 nm) are combined and transmitted over the same fiber cable. Multi-terabit capacities over a single fiber was achievable with use of this DWDM technology and it is likely that the next generation DWDM systems will require 40 Gbps or higher channel bit rate \cite{c4,c5}. Furthermore, multilevel modulation formats with single or multicarrier, polarization multiplexing, new detection schemes, novel fiber and Raman amplification technologies have been extensively investigated for higher data rates transmission up to 500 Gbps and 1 Tbps per channel \cite{c6}. Multilevel modulation formats with coherent detection is spectrally efficient, however it requires more energy for transmission than using simple modulation format with direct detection.
It was reported that transmission of 40 Gbps by 5 channels, with non return to zero – on / off keying (NRZ-OOK) modulation was power efficient after multiple fiber spans, as compared to dual polarization quadrature phase shift keying (DP-QPSK) modulation, in which, more than 90 \% of the total power consumption was due to re-amplification, re-shaping and re-timing for propagation distance more than 600 km \cite{c7}. The NRZ-OOK modulation format with direct detection is generally less expensive, because the transmitter and receiver configuration are simple to implement and the obtainable data rate and transmission distance may be sufficient in some propagation scenario, such as, for fiber connections from the mobile national core network to external gateway, core network to base station subsystems as well as within the base station subsystems. It is shown here in this work that it is possible to transmit up to 1.28 Tbps (Tera bits per second) using NRZ-OOK and direct detection, with 40 Gbps on each wavelength channel, for 32 channels, over 1000 km fiber span, which can be cost effective design for long haul mobile backbone in a national network.

\section*{Related works}
The modulation formats of carrier suppressed return-to-zero (CSRZ), duo binary return-to-zero (DRZ), and modified duo binary return-to-zero (MDRZ) were analyzed using pre, post and symmetrical dispersion compensation schemes with respect to Q
-	value and eye opening penalty for different transmission distances and signal input powers varying from -15 to10 dBm \cite{c8}. It was concluded that faithful transmission of 
32 channels by 40 Gbps (1.28 Tbps) over 1450 km was possible using MDRZ modulation format and symmetrical dispersion compensation scheme. However, MDRZ modulation format usually has a complex transmitter and receiver configuration \cite{c8}. 
	The performance analysis of DWDM System for different modulation schemes with varying channel spacing was reported in \cite{c9}. For 40 Gbps by 16 DWDM channels or 40 Gbps by 32 DWDM channels, it was reported that CSRZ scheme with 100 GHz channel spacing could reach transmission coverage up to 4,000 km for 16 and 32 channels DWDM system, because it is highly tolerant to nonlinear effects. MDRZ scheme with 75 GHz channel spacing exhibits transmission coverage up to 4,000 km and 4,500 km for 16 and 32 channels respectively, but with degraded signal due to effect of inter-symbol interference (ISI) \cite{c9}. However, many of such advanced modulation scheme is complex to implement at the transmitter and receiver side \cite{c8}. NRZ-OOK scheme seems to be the most simple to implement.
	The analysis and compensation of polarization mode dispersion (PMD) in single channel and 32 - channel DWDM system was reported in \cite{c10}. The simulation was carried out for 4-channel (40 Gbps), 8-channel (80 Gbps), 16-channel (160 Gbps) WDM systems and 32-channel (320 Gbps) DWDM fiber optic system with each channel having the data rate of 10 Gbps. The work focused on mitigation of polarization mode dispersion by deterministic differential group delay (DDGD) method for single channel and polarization maintaining fiber (PM) fiber method for multichannel compensation. The DWDM system with capacity up to 10 Gbps by 32 channels, 10 Gbps by 16 channels and 100 Gbps by 16 channels were simulated only over 100 km fiber span \cite{c10}.
	The simulation of DWDM systems with total capacity up to 1.28 Tbps over five spans of 50 km fiber (250 km length) and spectral efficiency approaching 0.4 bps/Hz was reported \cite{c11}. The impact of signal-to-noise ratio on parameters such as channel spacing, fiber length, dispersion, and number of channels were investigated \cite{c11}. The signal to noise ratio improved as the channel spacing was increased. However, in addition to fiber span increase, there is need to improve the spectra efficiency, and as such the channel spacing needs to be optimal.
Hoshida et al. \cite{c3} presented a performance comparison of modulation formats of non return-to-zero (NRZ), CSRZ, and bit-synchronous intensity modulated differential phase shift keying (IM-DPSK) format in 75-GHz (0.6 nm) spaced WDM long-haul transmission systems. It was concluded that NRZ format is good for shorter transmissions distances up to 1000 km, and therefore is attractive with its virtues such as simple and low-cost transmitter and receiver configuration with small dependence on fiber type in terms of nonlinear tolerance. CS-RZ format is less attractive in highly spectral efficient systems with less than 75-GHz (0.6 nm) spacing. IM-DPSK was concluded to be the best choice among the three for transmission distance beyond 1000 km \cite{c3}. 
Here in this work, a 40 Gbps by 32 channel DWDM system is designed and simulated using channel spacing of 0.4 nm (50 GHz) corresponding to a spectra efficiency of 0.8 bps/Hz. The NRZ - OOK simple modulation format is used to achieve a faithful transmission distance reaching 1000 km.

%\begin{document}
  \begin{figure*}[t]
      \centering
			\includegraphics[width=\textwidth]{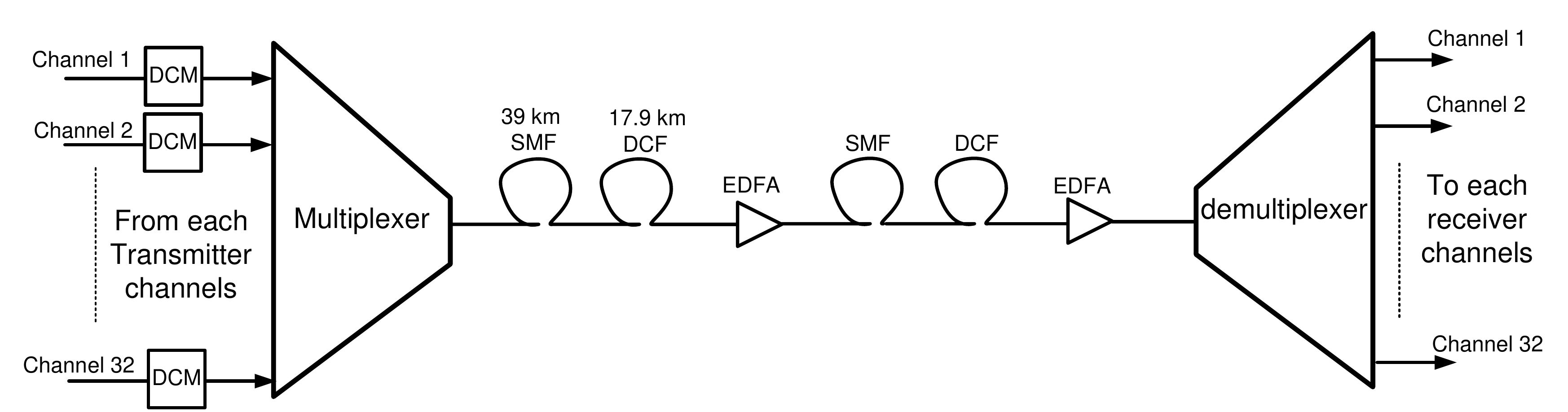}
      \caption{Illustration of the simulated DWDM system}
      \label{figurelabel1}
   \end{figure*}
	%\end{document}
\section*{Methodology}
RSoft OptSim$^{TM}$ software from Synopsys Inc is used for the system simulation. The software solves the nonlinear Schrödinger equation using a time domain split-step algorithm \cite{c7}. The software contains models for different optical devices and the analyzing tools, such as the laser, external modulator, multiplexer / demultiplexer, single mode fiber (SMF), dispersion compensation fiber (DCF), dispersion compensated module (DCM), erbium doped fiber amplifier (EDFA), optical filters, optical receiver (photodetector), the spectrum analyzer, power meter, bit error rate tester, eye diagram analyzer, and property map. The software accuracy depends on the number of the simulated bits, in which, the Q-factor uncertainty (error) is usually less than 0.28 dB for at least 8000 bits \cite{c7}.
At first, the transmission limit for single channel 40 Gbps data rate in a single mode fiber with fiber loss of 0.2 dB/km is obtained by setting the dispersion and non-linearity parameters to be zero and obtaining a faithful transmission with appreciable eye opening and bit error rate less than 10$^{-9}$. This transmission limit was determined to be about 55 km, which corresponds to the amplifier spacing of the system.
The dispersion limit of the single mode fiber was obtained to be about 8 km, which was determined by setting the dispersion parameter to regular value for single mode fiber (18 ps / nm-km). Material dispersion is a property of the fiber in which the different wavelength components propagates at different speed, which could lead to inter–symbol interference. Dispersion compensation fiber is therefore necessary in the system for long haul transmission, since the system is greatly dispersion limited. Corning SMF 1550 is used as the single mode fiber, while Corning Vascade S – 1000 is used as the dispersion compensating fiber. The required amount of dispersion compensation is determined by the following equation \cite{c12}:
\begin {equation} 
({{D}}({\lambda}_{SMF})\times{L}_{SMF})+({D}({\lambda}_{DCF})\times{L}_{DCF}) = 0 
\end {equation} 			
Where, D(${\lambda}_{SMF}$) and D(${\lambda}_{DCF}$) are the dispersion parameters in ps / (nm-km) for single mode fiber (SMF) and dispersion compensation fiber (DCF) respectively, while L$_{SMF}$ and L$_{DCF}$ are lengths of the SMF and DCF respectively. The SMF has a positive dispersion parameter (+18 ps / (nm-km) for Corning SMF 1550), which is compensated by DCF with large negative dispersion parameter (-38 ps / (nm-km) for Corning Vascade S – 1000).
 
\begin{figure}[h!]
      \centering
			\includegraphics[width=0.5\textwidth]{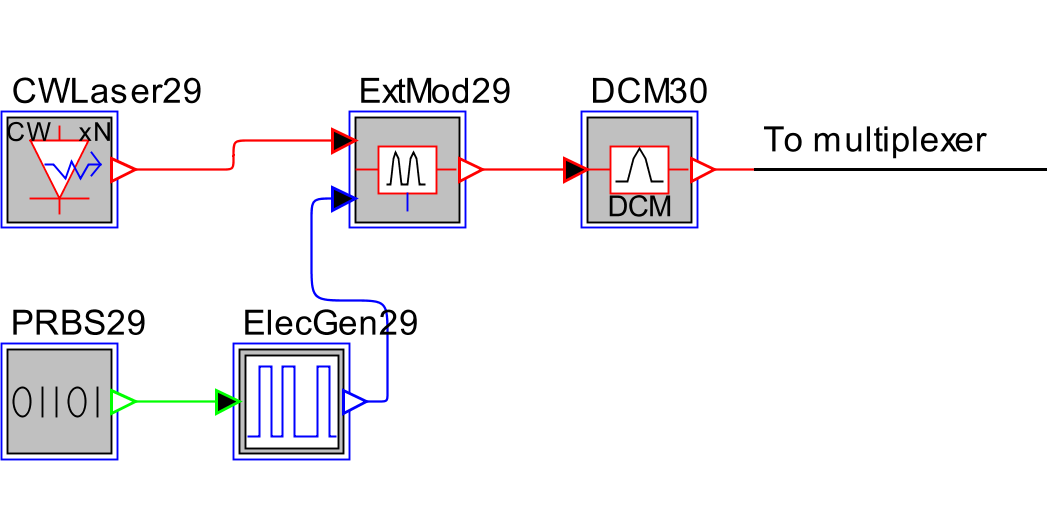}
      \caption{Single channel transmitter}
      \label{figurelabel2}
   \end{figure}

The simulation is then extended to a DWDM system consisting of 32 wavelength channels in which each wavelength is modulated at a standard data rate of 40 Gbps. The wavelength range is chosen from 1.5436 to 1.556 nm which falls within the C – band range, while the channel spacing was chosen to be 0.4 nm. The wavelength spacing corresponds to a frequency spacing of 50 GHz which gives a spectra efficiency of 0.8 bps/Hz for 40 Gbps. The simulated system is shown in Figure \ref{figurelabel1}. This starts from the transmitters for each channels, to multiplexer, single mode fiber (SMF), dispersion compensation fiber (DCF), erbium doped fiber amplifier (EDFA), demultiplexer and then to optical receivers for each channels. Detailed description of each channel is shown in Figure \ref{figurelabel2}.

	%\begin{document}
\begin{figure}[h!]
      \centering
			\includegraphics[width=0.5\textwidth]{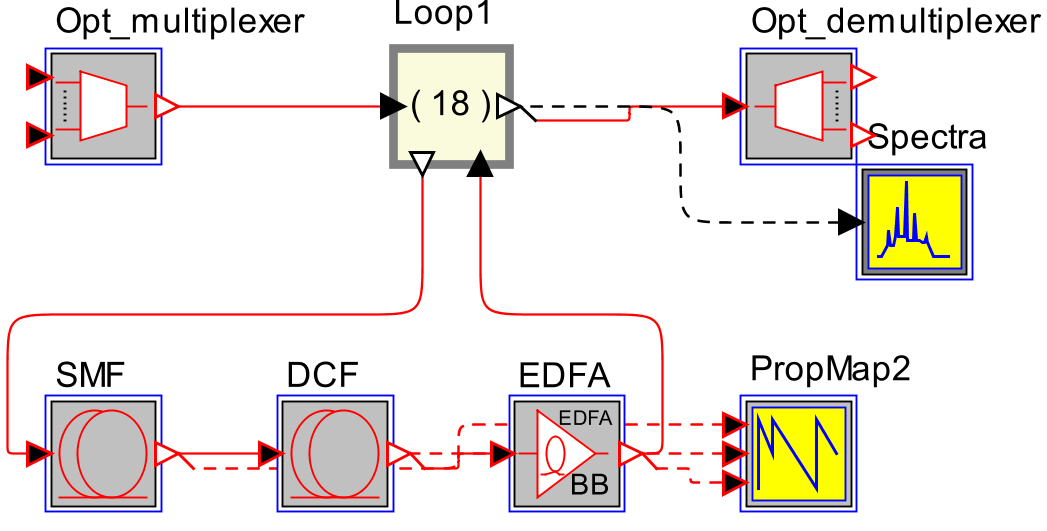}
      \caption{Transmission link showing SMF, DCF and EDF}
      \label{figurelabel3}
   \end{figure}

	%\end{document}
A pseudo random bit sequence (PBRS) at data rate of 40 Gbps is sent to the electrical generator. The output of the electrical generator is subsequently used to modulate the continuous wave laser using the external modulator. A pre–dispersion compensation module (DCM) is used to pre–compensate for material dispersion that will occur in the fiber and this gives a modulated optical output for each channel. This output is then sent to the multiplexer. The multiplexer (Figure \ref{figurelabel3}) combines the modulated output from each channel (32 channels) to obtain a cumulative data rate of 1.28 Tbps and the output is transmitted over the fiber.

\begin{figure}[h!]
      \centering
			\includegraphics[width=0.55\textwidth]{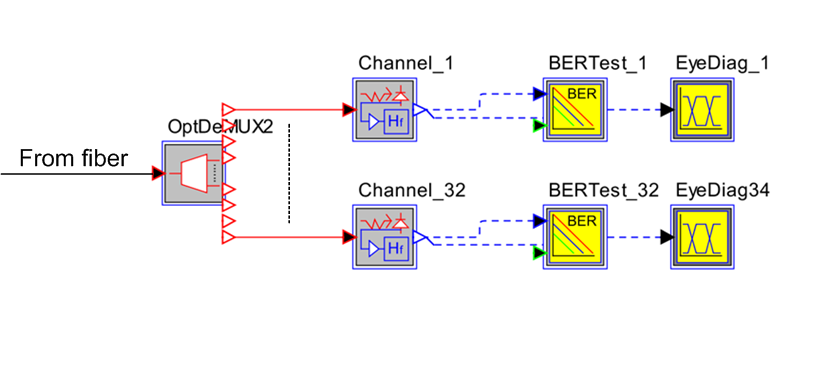}
      \caption{Receiver section showing the multiplexers and the photodetectors}
      \label{figurelabel4}
   \end{figure}

A loop is used to combine the SMF, inline DCF and EDFA until the transmission distance of 1000 km is reached. A total of 18 loops of SMF, DCF and EDFA are required to reach transmission distance of 1000 km.
The receiver section is shown in Figure \ref{figurelabel4}. The output from the fiber is connected to a demultiplexer at the receiving end. This separates the multiplexed wavelength channels into individual channels and the optical signals are directly detected using photo detectors.

\section*{Results and Discussions}
\begin{figure}[h!]
      \centering
			\includegraphics[width=0.5\textwidth]{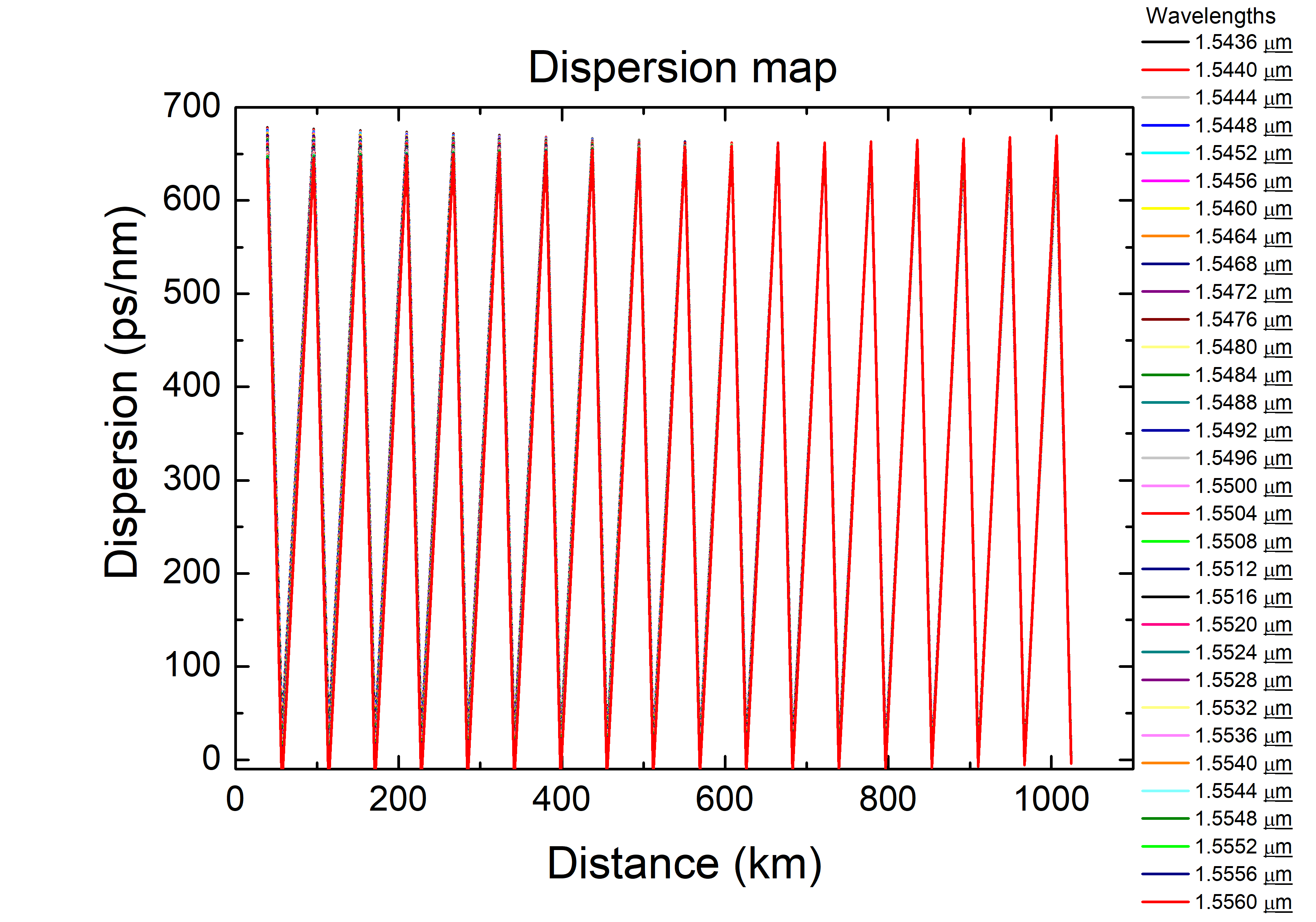}
      \caption{Measured dispersion with respect to fiber distance }
      \label{figurelabel5}
   \end{figure}

Careful dispersion compensation was done to obtain appreciable eye-opening and bit error rate less than 10$^{-9}$ in all the channels for 1000 km transmission distance. A pre-dispersion compensation module (DCM) was also used in all the channels to compensate for the final dispersion measured with the dispersion map at the receiver end. The dispersion for SMF length of 39 km was therefore balanced with DCF length of 17.9 km resulting to total fiber lengths of 56.9 km (amplifier spacing). The measured dispersion with respect to fiber lengths is shown in Figure \ref{figurelabel5}.

\begin{figure}[h!]
      \centering
			\includegraphics[width=0.5\textwidth]{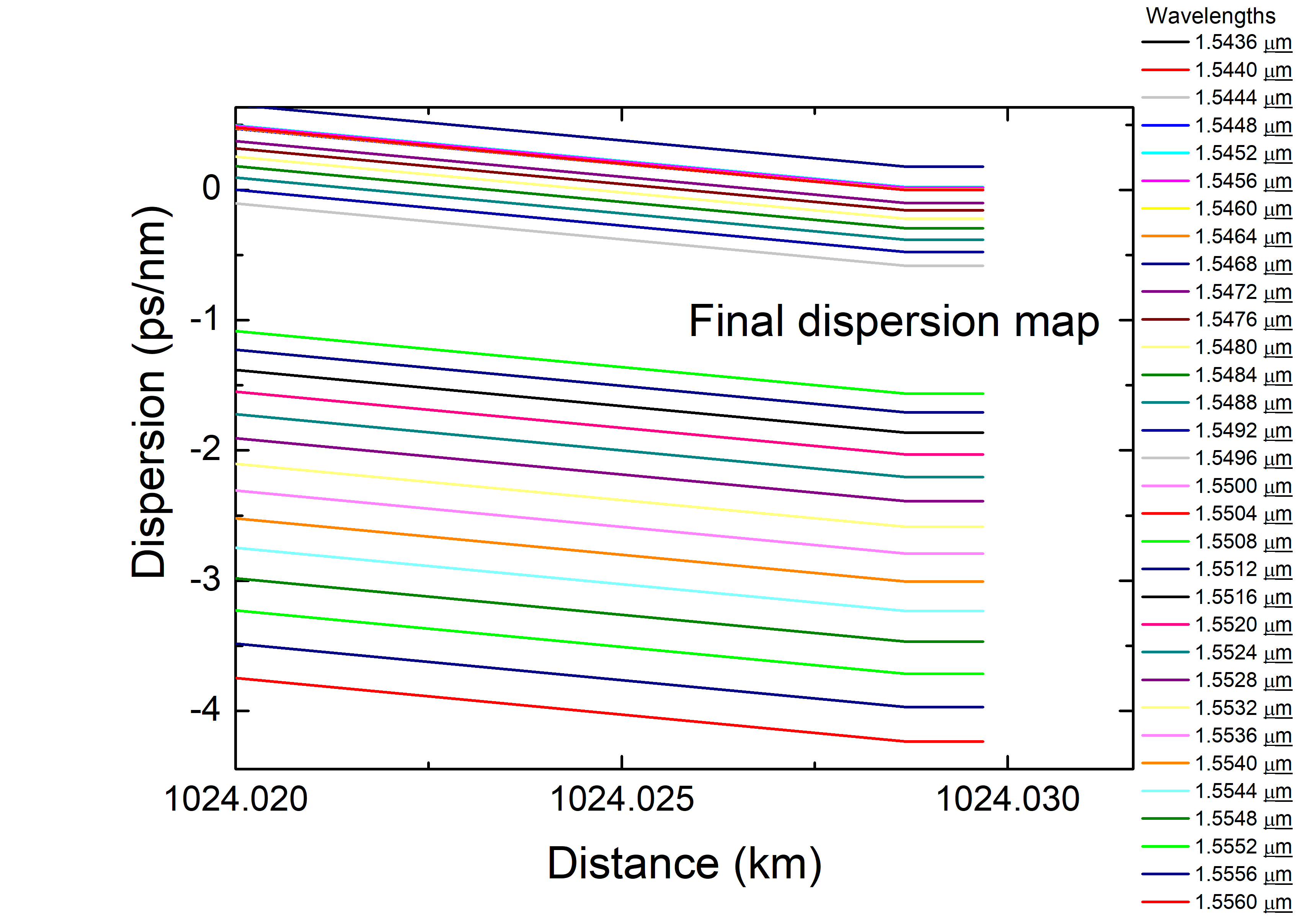}
      \caption{Final measured dispersion at 1000 km distance}
      \label{figurelabel6}
   \end{figure}

The dispersion varied from a large positive value (due to SMF dispersion) and decreased linearly as a result of compensation with DCF. The final dispersion at 1000 km is shown in Figure \ref{figurelabel6}. 
	
\begin{figure}[h!]
\centering
   \begin{subfigure}[b]{0.5\textwidth}
   \includegraphics[width=\textwidth]{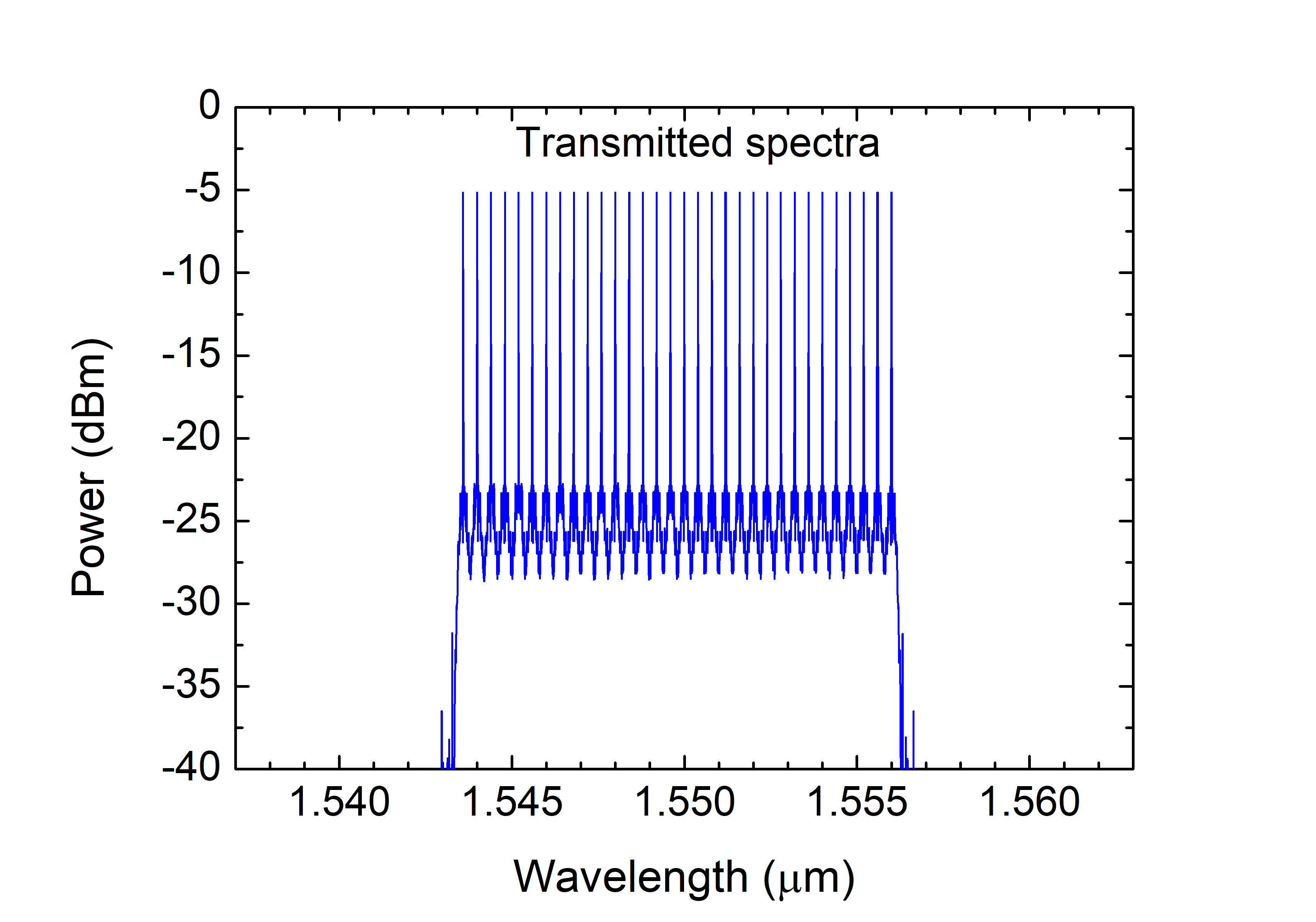}
   \caption{Transmitted spectrum}
   \label{figurelabel7} 
\end{subfigure}
\begin{subfigure}[b]{0.5\textwidth}
   \includegraphics[width=\textwidth]{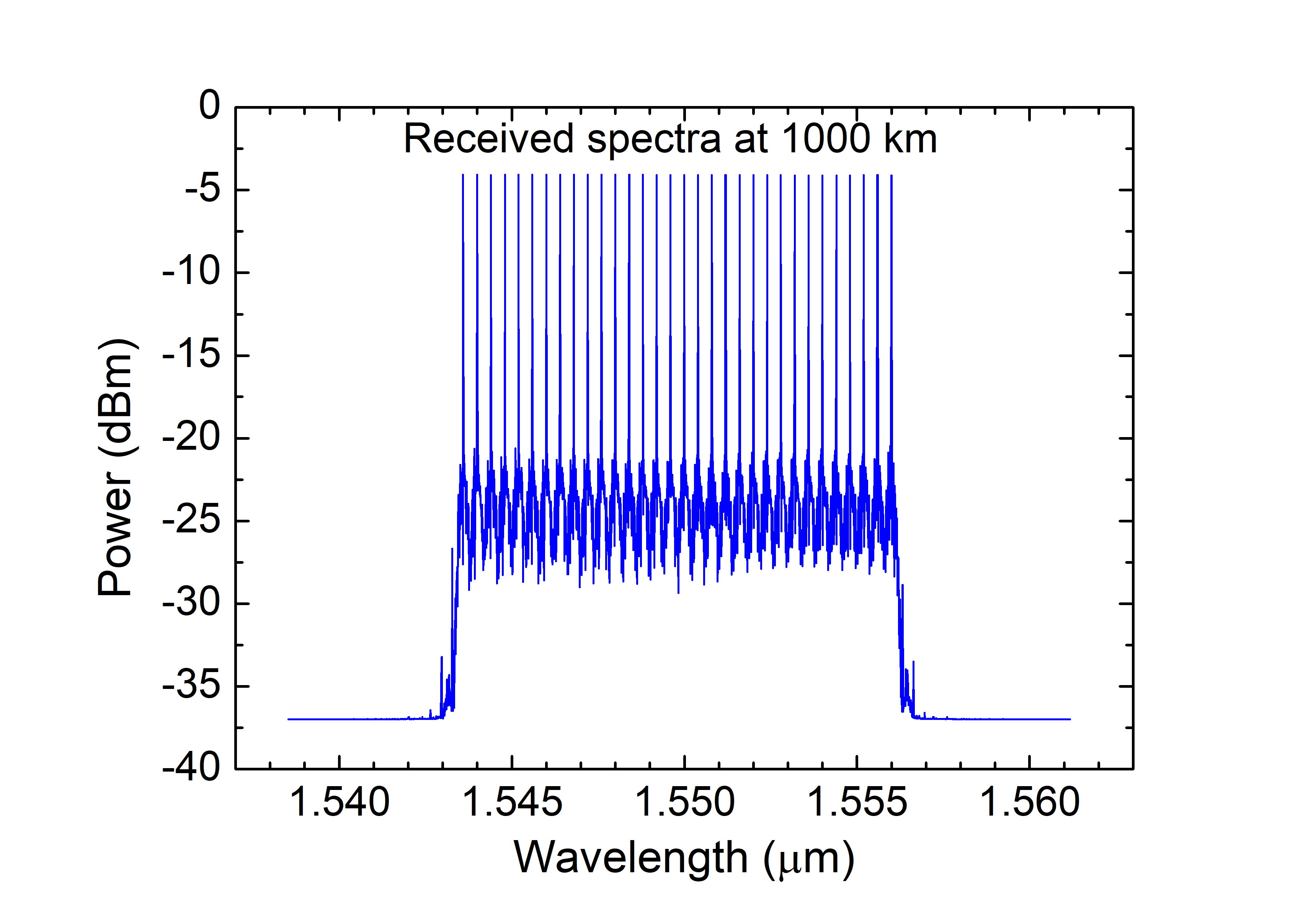}
   \caption{Received spectrum}
   \label{figurelabel7b}
\end{subfigure}
\caption{Transmitted and the received spectra at 1000 km for the 32 channels}
\end{figure}

In order to obtain an appreciable eye opening and bit error rate less than 10$^{-9}$ in all channels, the final dispersion varies between -4 to 0.6 ps-nm. The system could tolerate more negative dispersion than positive dispersion. In addition, the received spectrum is a replica of the transmitted spectrum (as shown in Figures \ref{figurelabel7} and \ref{figurelabel7b}).

\begin{figure}[h!]
      \centering
			\includegraphics[width=0.5\textwidth]{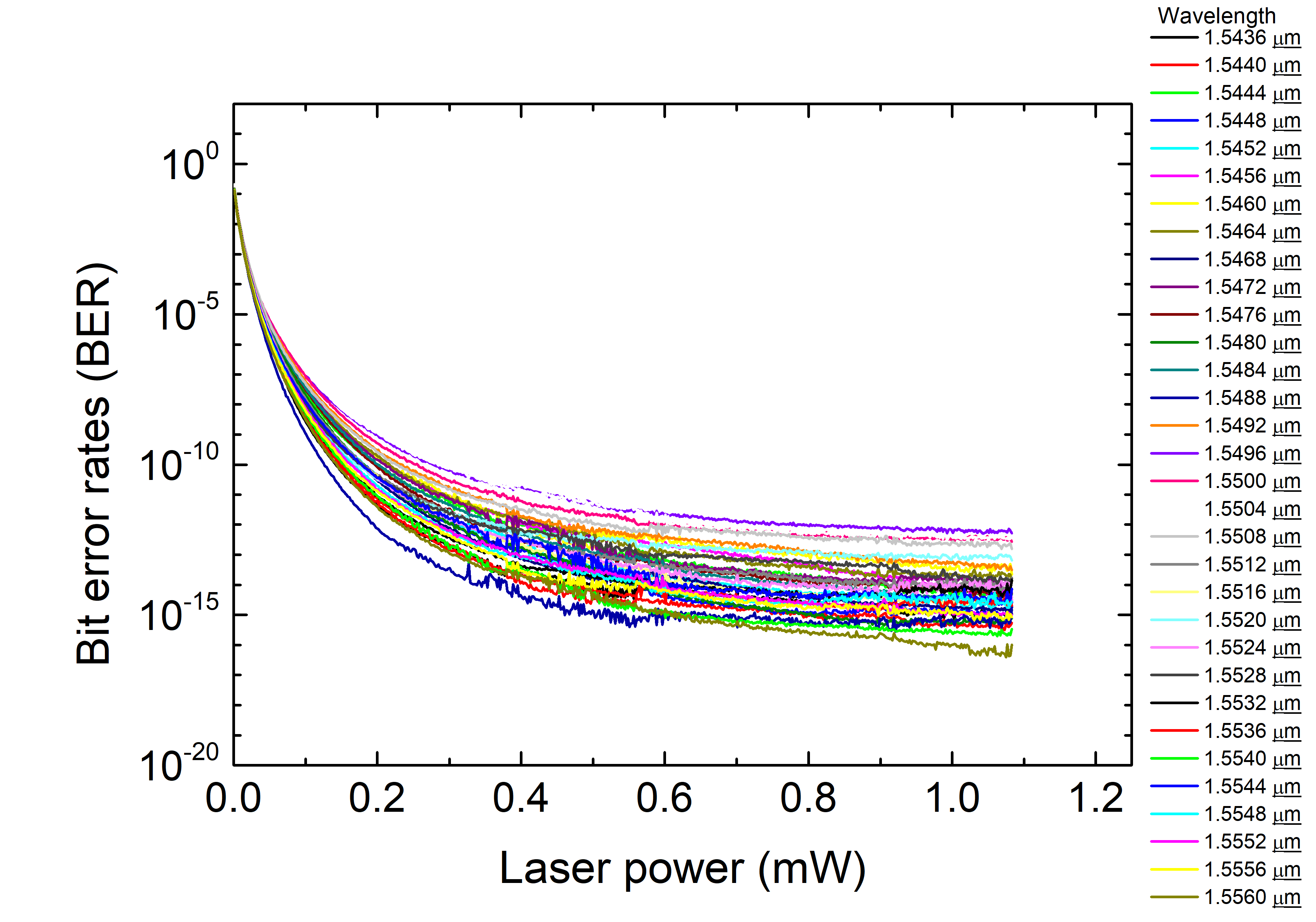}
      \caption{Received BER as a function of laser transmitter power}
      \label{figurelabel8}
   \end{figure}

This is an indication of low bit error rates in all the channels, if the received spectrum is a replica of transmitted spectrum, since inter channel cross talk is avoided after careful dispersion compensation.
The received bit error rates (BER) in all channels as a function of the laser power (transmitter) is shown in Figure \ref{figurelabel8}. 

\begin{figure}[h!]
      \centering
			\includegraphics[width=0.5\textwidth]{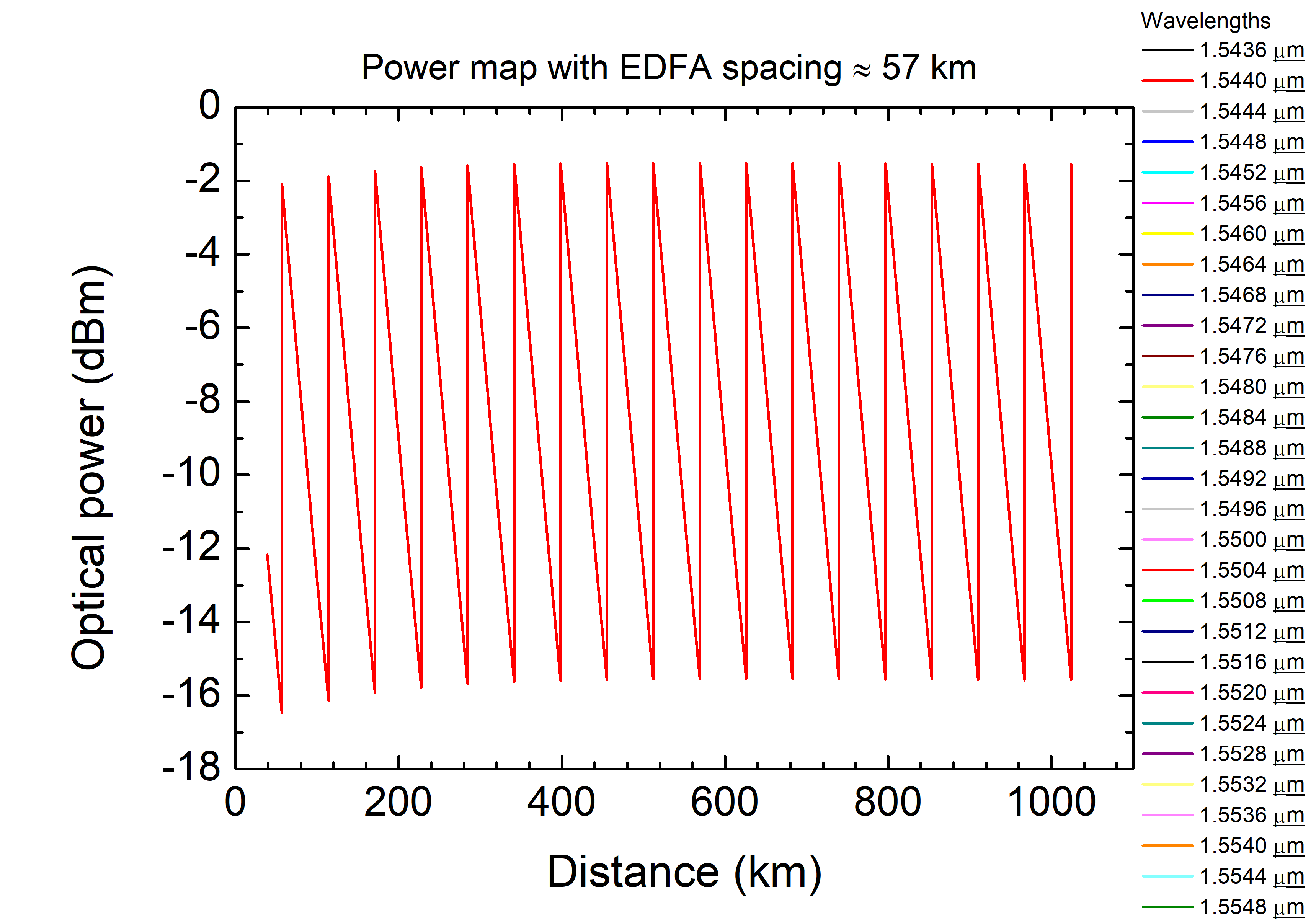}
      \caption{Optical transmitter power as a function of fiber length}
      \label{figurelabel9}
   \end{figure}

It can be observed that the BER decreased as the laser power is increased up to an optimal power level, where the BER becomes independent of the laser power increase. Transmitting too high power is usually detrimental to the system since this can induce noise in the received spectra and inter channel cross talk due to fiber non - linearity \cite{c13}. 
The transmitted optical power as a function of fiber length is shown in Figure \ref{figurelabel9}. 

\begin{figure}[h!]
      \centering
			\includegraphics[width=0.5\textwidth]{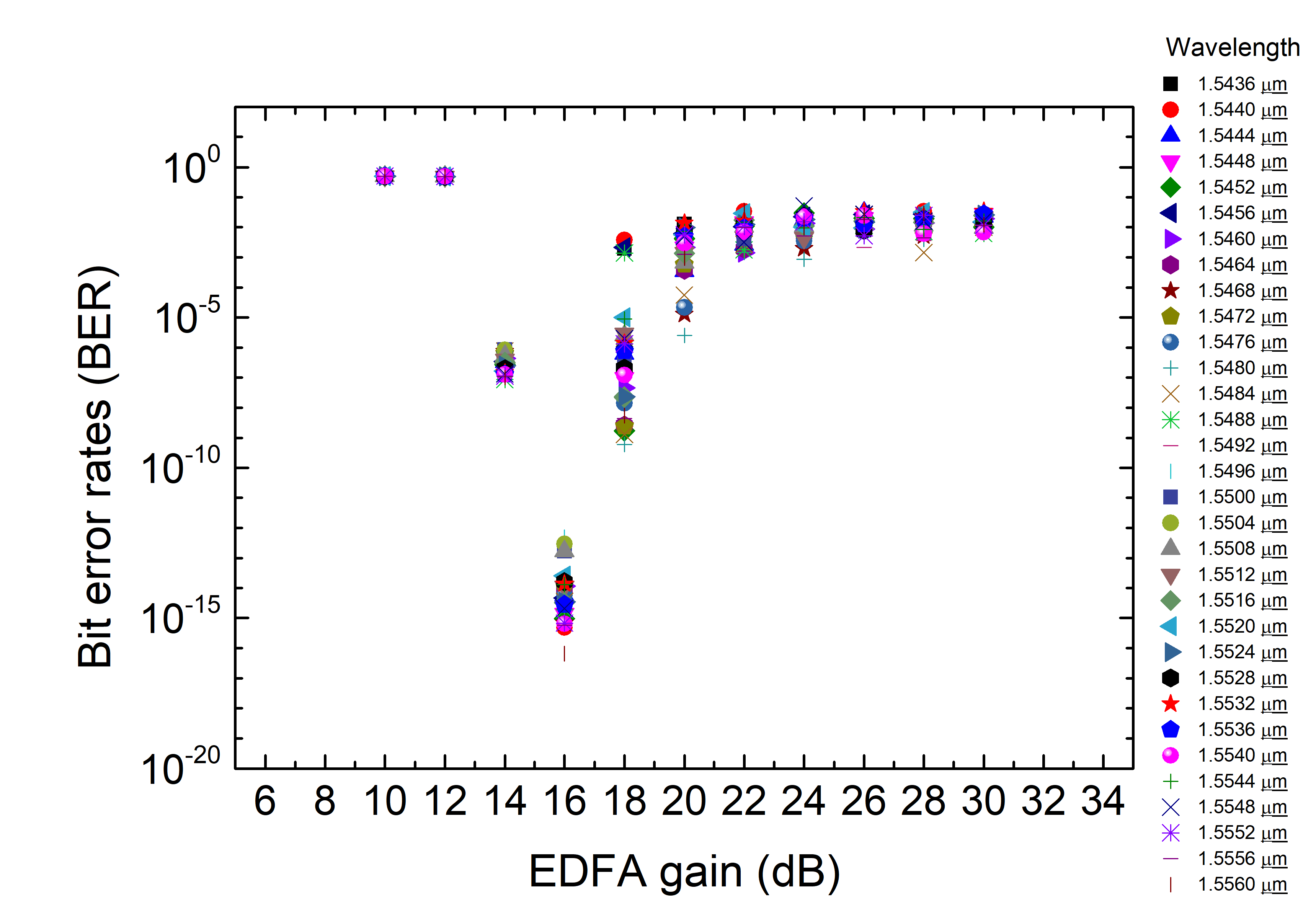}
      \caption{Received BER as a function of EDFA gain}
      \label{figurelabel10}
   \end{figure}

A -12 dBm laser transmitter power was found to be optimum in all channels to obtain a BER less than 10$^{-9}$ in the channels. A decrease in the optical power along the line was due to fiber loss, in which erbium doped fiber amplifier (EDFA) is used to periodically amplify the signal at EDFA spacing of approximately 57 km. 
The received BER as a function of EDFA gain is shown in Figure \ref{figurelabel10}.  

The optimum EDFA gain was -16dB. High BER was obtained for over 18 dB EDFA gain due to stimulation of fiber non-linearity from high optical power. Another critical criterion to obtain a faithful transmission in all the channels is the choice of WDM demultiplexer filter parameters, such as the demultiplexer filter bandwidth and the demultiplexer filter spacing. Tight optical filtering is necessary at the transmitter and the receiver side, in order to limit inter-channel interference \cite{c14}.

 \begin{figure}[h!]
      \centering
			\includegraphics[width=0.5\textwidth]{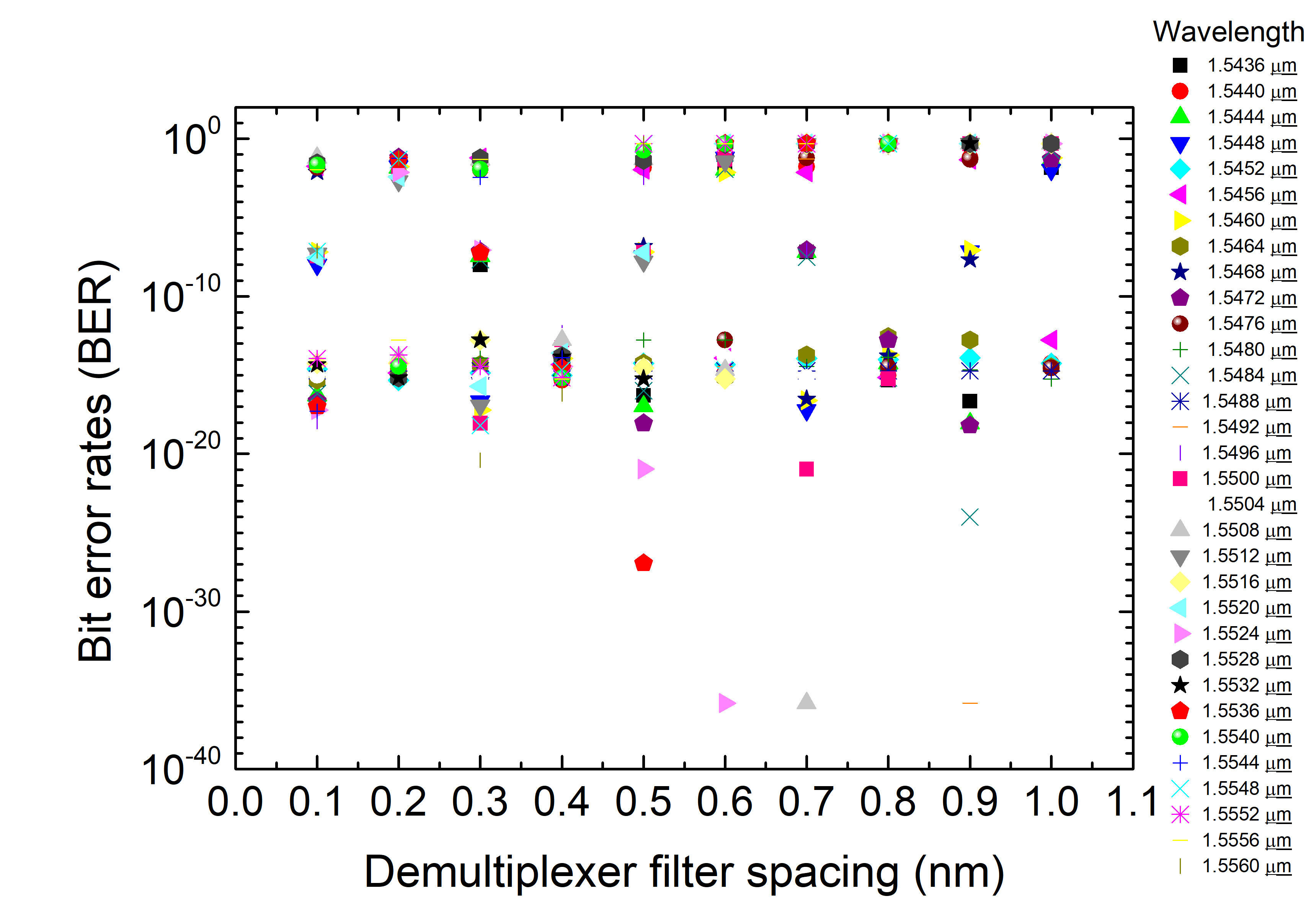}
      \caption{Received BER as a function of demultiplexer spacing}
      \label{figurelabel11}
   \end{figure}

\begin{figure}[h!]
      \centering
			\includegraphics[width=0.5\textwidth]{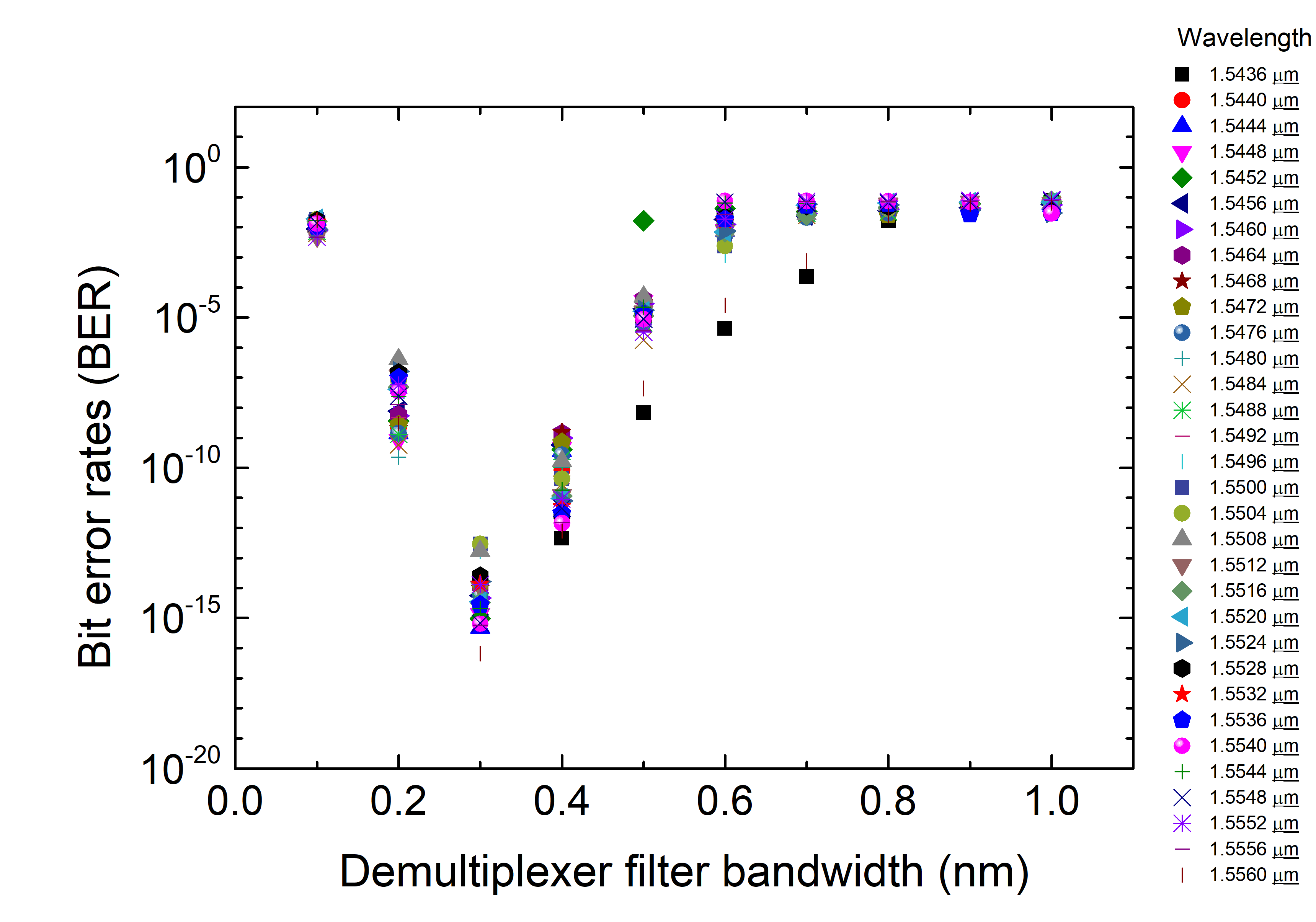}
      \caption{Received BER as a function of demultiplexer bandwidth}
      \label{figurelabel12}
   \end{figure}

	The demultiplexer filter spacing separates the multiplexed signal into individual wavelength channels, while the demultiplexer filter bandwidth (full width half maximum bandwidth) selects the modulated data along with each wavelength channel. The BER as a function of demultiplexer filter spacing and filter bandwidth are shown in Figures \ref{figurelabel11} and \ref{figurelabel12} respectively.
	
\begin{figure}[h!]
\centering
   \begin{subfigure}[b]{0.5\textwidth}
   \includegraphics[width=\textwidth]{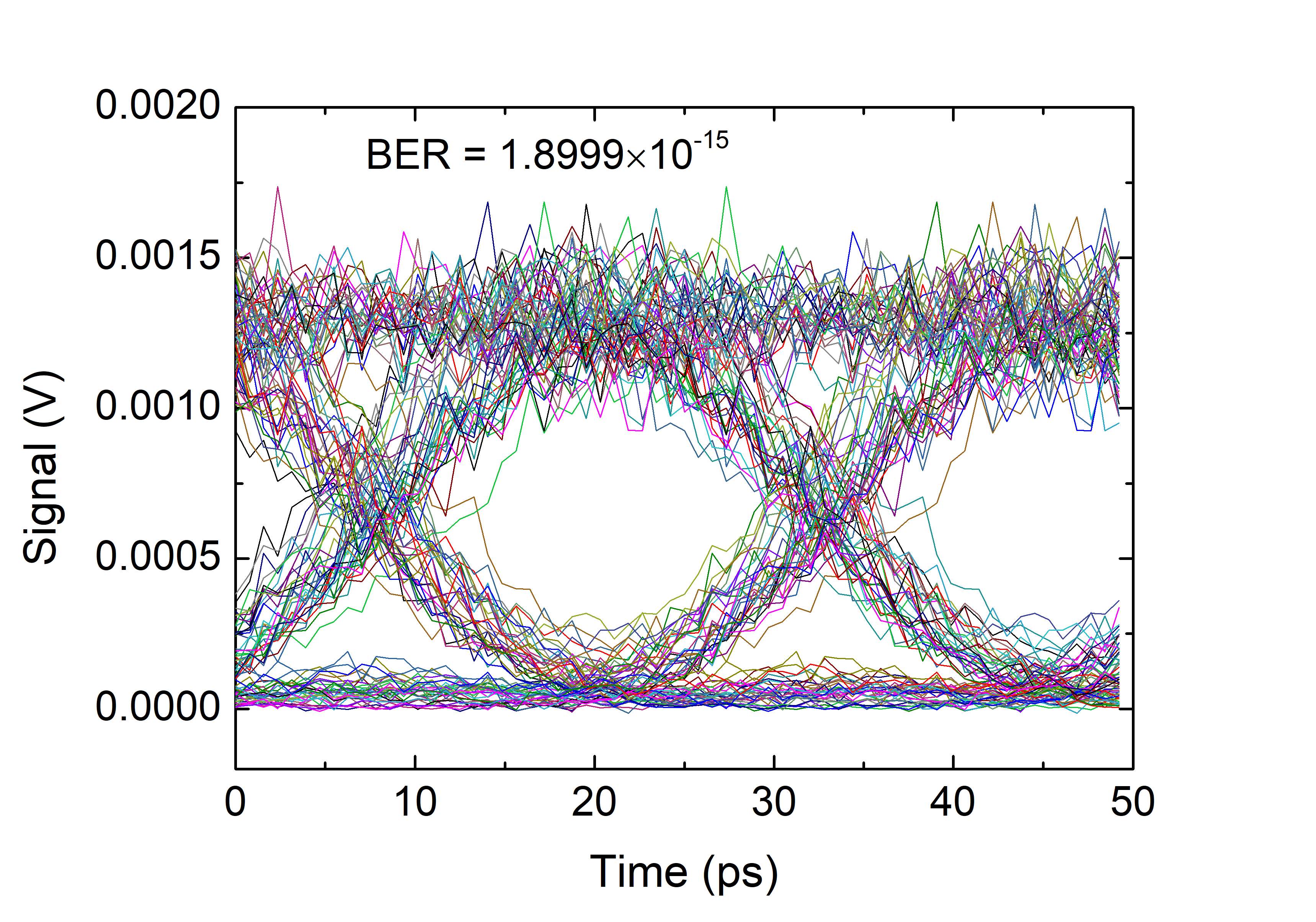}
   \caption{Channel 1}
   \label{figurelabel13} 
\end{subfigure}
\begin{subfigure}[b]{0.5\textwidth}
   \includegraphics[width=\textwidth]{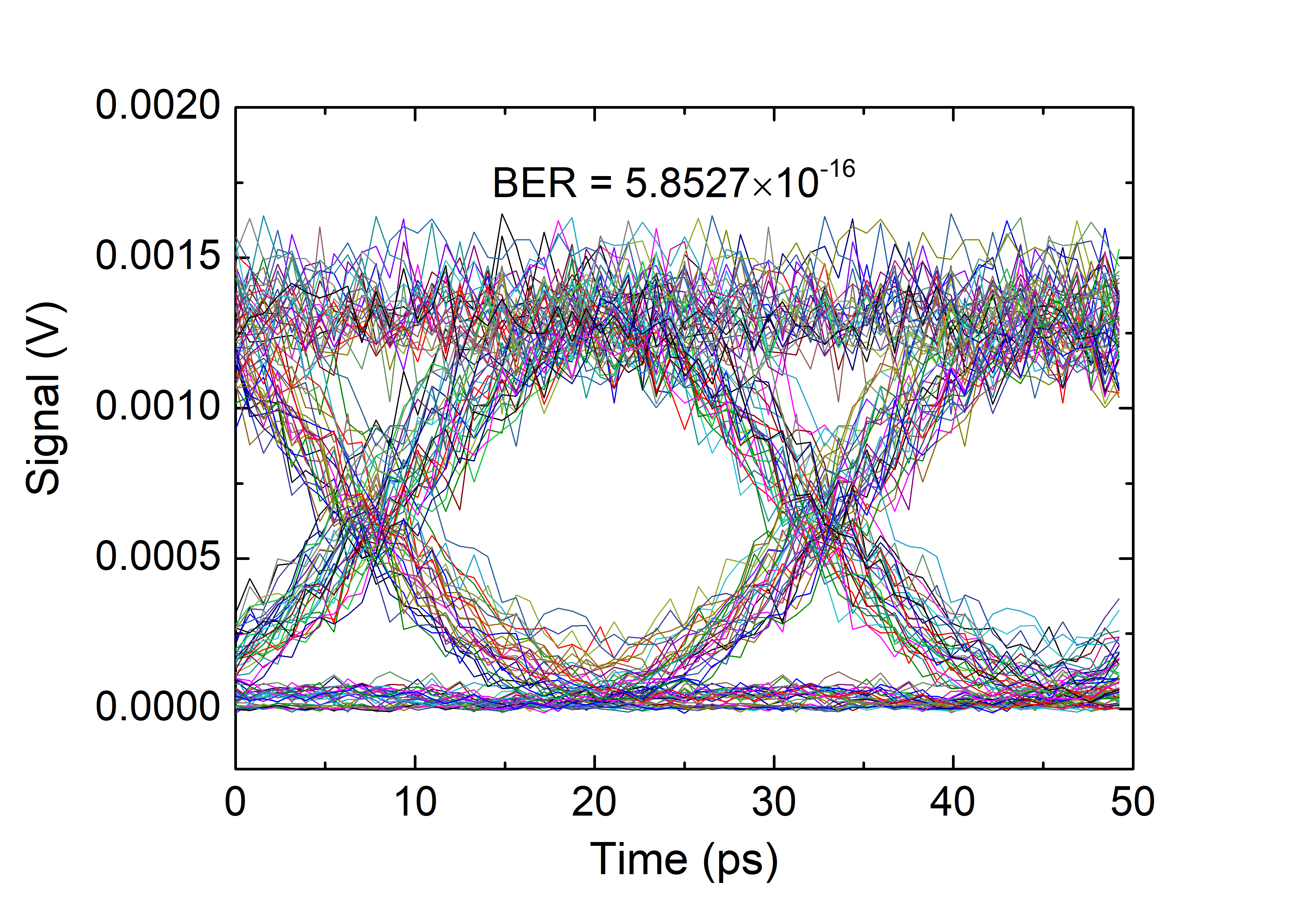}
   \caption{Channel 32}
   \label{figurelabel13b}
\end{subfigure}
\caption{Optimized eye diagram for channel 1 and channel 32}
\end{figure}

The demultiplexer filter spacing is chosen to be equal to the channel spacing used for the DWDM. Choosing too large filter spacing will lead to inter-band cross talk in some channels and too low spacing will result to insufficient filtering (cutoff) in some channels. It can be observed that the BER was below 10$^{-10}$ for the entire wavelengths at 0.4 nm filter spacing which corresponds to the given DWDM channel spacing. In addition, all the wavelength channels gave BER below 10$^{-10}$ at 0.3 nm filter bandwidth. 
The eye diagram openings were identical in all the channels and examples of the eye diagram opening for channel 1 and channel 32 are shown in Figures \ref{figurelabel13} and \ref{figurelabel13b}.
	
\section*{Conclusion}

A DWDM long haul transmission system up to 1000 km fiber length has been designed and simulated using commercial simulation software. The standard channel spacing of 0.4 nm was used, in the C-band range from 1.5436 - 1.556 nm, to transmit 40 Gbps data by 32 wavelength channels using simple modulation format of NRZ - OOK and a direct detection scheme. A faithful transmission with BER less than 10$^{-9}$ was obtained in all the channels, in which the critical criteria for the design were accurate dispersion management, optimal transmitter power, EDFA gain, as well as DWDM filter spacing and bandwidth. It is proposed that the design is a cost effective design for long haul backbone in a national network for a mobile communication system, since NRZ - OOK and a direct detection scheme is simple and cheaper than advanced modulation schemes and coherent detection. A total bit rate length product of 1.28 Pbps.km was achievable.


\begin{thebibliography}{99}

\bibitem{c1} Agrell, E., Karlsson, M., Chraplyvy, A. R., Richardson, D. J., Krummrich, P.M., Winzer, P., Roberts, K., Fischer, J.K., Savory, S.J., Eggleton, B.J. and Secondini, M., (2016). Roadmap of optical communications. Journal of Optics, 18(6), 063002.
\bibitem{c2} Liu, G., Jiang, D. (2016). 5G: Vision and requirements for mobile communication system towards year 2020. Chinese Journal of Engineering, 2016.
\bibitem{c3} Hoshida, T., Vassilieva, O., Yamada, K., Choudhary, S., Pecqueur, R., and Kuwahara, H. (2002). Optimal 40 Gb/s modulation formats for spectrally efficient long-haul DWDM systems. Journal of lightwave technology, 20(12), 1989-1996.
\bibitem{c4} Nielsen, T. N. (2000). 3.28-Tbit/s (82× 40 Gb/s) transmission over 3×100 km nonzero-dispersion fiber using dual C-and L-band hybrid Raman/Erbium-doped inline amplifiers. Technical Digest of OFC2000, postdeadline paper.
\bibitem{c5} Bonati, A., Chesnoy, J., Erman, M., Gabla, P. M., Piacentini, B., and Reinaudo, C. (1999). Global turnkey solutions for backbone transmission networks. Alcatel telecommunications review, (3), 205-218.
\bibitem{c6} Yu, J., Zhou, X., (2010) Ultra-high-capacity DWDM transmission system for 100G and beyond. IEEE Communications Magazine. 48(3).
\bibitem{c7}Udalcovs, A., Monti, P., Bobrovs, V., Schatz, R., and Wosinska, L. (2014). Power efficiency of WDM networks using various modulation formats with spectral efficiency limited by linear crosstalk. Optics Communications, 318, 31-36.
\bibitem{c8} Sheetal, A., Sharma, A. K., and Kaler, R. S. (2010). Simulation of high capacity 40 Gb/s long haul DWDM system using different modulation formats and dispersion compensation schemes in the presence of Kerr's effect. Optik-International Journal for Light and Electron Optics, 121(8), 739-749.
\bibitem{c9} Sharma, D., and Prajapati, Y. K. (2016). Performance analysis of DWDM system for different modulation schemes using variations in channel spacing. Journal of Optical Communications, 37(4), 401-413.
\bibitem{c10} Sabapathi, T., and Manohari, R. G. (2014). Analysis and compensation of polarization mode dispersion in single channel, WDM and 32-channel DWDM fiber optic system. Optik-International Journal for Light and Electron Optics, 125(1), 18-24.
\bibitem{c11} Kaler, R. S., Kamal, T. S., and Sharma, A. K. (2002). Simulation results for DWDM systems with ultra-high capacity. Fiber and Integrated Optics, 21(5), 361-369.
\bibitem{c12} Alsevska, A., Dilendorfs, V., Spolitis, S., and Bobrovs, V. (2017). Comparison of Chromatic Dispersion Compensation Method Efficiency for 10 Gbit/S RZ-OOK and NRZ-OOK WDM-PON Transmission Systems. Latvian Journal of Physics and Technical Sciences, 54(6), 65-75.
\bibitem{c13} Singh, S. P., and Singh, N. (2007). Nonlinear effects in optical fibers: Origin, management and applications. Progress in Electromagnetics Research, 73, 249-275.
\bibitem{c14} Bosco, G., Carena, A., Curri, V., Gaudino, R., and Poggiolini, P. (2004). Modulation formats suitable for ultrahigh spectral efficient WDM systems. IEEE journal of selected topics in quantum electronics, 10(2), 321-328.


\end{thebibliography}
\end{document}